# Optical coupling between atomically-thin black phosphorus and a two dimensional photonic crystal nanocavity

Yasutomo Ota[1], Rai Moriya[2], Naoto Yabuki[2], Miho Arai[2], Masahiro Kakuda[1],
Satoshi Iwamoto[1,2], Tomoki Machida[1,2] and Yasuhiko Arakawa[1,2]
E-mail: ota@iis.u-tokyo.ac.jp

1) Institute for Nano Quantum Information Electronics, The University of Tokyo, 4-6-1 Komaba, Meguro-ku, Tokyo 153-8505, Japan

2) Institute of Industrial Science, The University of Tokyo, 4-6-1 Komaba, Meguro-ku, Tokyo 153-8505, Japan

Atomically-thin black phosphorus (BP) is an emerging two dimensional (2D) material exhibiting bright photoluminescence in the near infrared. Coupling its radiation to photonic nanostructures will be an important step toward the realization of 2D material based nanophotonic devices that operate efficiently in the near infrared, which includes the technologically important optical telecommunication wavelength bands. In this letter, we demonstrate the optical coupling between atomically-thin BP and a 2D photonic crystal nanocavity. We employed a home-build dry transfer apparatus for placing a thin BP flake on the surface of the nanocavity. Their optical coupling was analyzed through measuring cavity mode emission under optical carrier injection at room temperature.

Nanophotonics employing two dimensional (2D) materials has been under intensive study with prospects for a wide range of applications such as optical modulators[1–3], detectors[4–6], nonlinear optical devices[7,8] and light sources[9–13]. Regarding the development of light sources, the use of 2D materials with a direct bandgap is highly advantageous due to their stronger radiative carrier recombination than those with an indirect bandgap. There are several 2D materials exhibiting bright luminescence with a direct bandgap, such as monolayer transition metal dichalcogenides[14]. Coupling them to photonic nanosturctures has also been sought after for the realization of efficient light sources[10–12,15]. However, their emission wavelength bands are predominantly in/around the visible range, leaving the development of 2D material light sources in the near infrared (NIR) (which includes the technologically important telecommunication bands).

In this context, atomically-thin black phosphorus[16] (BP) is an exceptional 2D material due to its strong luminescence in the NIR originated from its direct bandgap nature[17–19]. In contrast to other 2D materials emitting in the NIR[13], BP keeps the bright luminescence even with increasing number of atomic layers, while shifting its emission wavelength from ~ 800 nm (monolayer) to >1600 nm (4 layers). Another unique optical property of atomically-thin BPs is its highly linearly polarized optical response[6,17–19]. All of these optical properties seem to be very useful for developing novel 2D material-based nanophotonic devices in the NIR.

A straightforward route for the development of nanophotonic light source based on atomically-thin BP is to combine it with optical nanocavities, such as those based on photonic crystals (PhCs)[10,13] and plasmonic structures[15], in which the light matter interaction can be significantly enhanced due to their tight optical confinement both in time and space. However, so far, there is no report on the optical coupling between atomically-thin BP and nanocavities.

In this letter, we report the observation of cavity-coupled NIR emission from a 2D BP flake directly placed on a 2D PhC nanocavity. We fabricated the structure using a dry transfer apparatus[20], which is equipped in an inert gas atmosphere to prevent the notoriously rapid degradation of BP flake in air. The fabricated BP-nanocavity system exhibits strong photoluminescence (PL) in the NIR, accompanied with a sharp cavity resonance peak due to the BP-nanocavity optical coupling. Our work here is an important step towards the development of BP based nanophotonic devices, including 2D material NIR nanolasers.

A schematic illustration of the BP-nanocavity coupled system investigated in this study is shown in Fig. 1(a). An atomically-thin BP flake is directly attached onto the surface of a 2D photonic crystal nanocavity. We fabricated this structure using a dry transfer technique based on that widely used in 2D material research[20]. The process consists of (1) the preparation of an ensemble of BP flakes on a soft elastomer stamp, (2) searching an appropriate thin BP by optical measurements and (3) transfer printing the selected BP flake onto the nanocavity surface under an optical microscope. In the following, we further describe each step.

First, we prepared thin BP flakes on an adhesive tape (SPV-224, Nitto) by repeated mechanical exfoliation, starting from a portion of bulk BP (Smart Elements). The prepared BP flakes are directly transferred to a polydimethylsiloxane (PDMS) film (Gel-Film x4, Gelpak) through putting and quickly pealing the tape embedding the flakes. Then, the elastomeric PDMS film with the transferred thin BPs is set in a vacuum chamber for subsequent optical measurements. An important point here is that we used a glove box to keep the BP flakes under dry nitrogen gas atmosphere during all the processes from the mechanical exfoliation to the sample setting. This is essential for preventing the notoriously rapid degradation of atomically-thin BP in air[21]. Indeed, we observed improvement of luminescence intensity of BP flakes on PDMS when we used purer nitrogen gas atmosphere in the glove box.

Second, we seek an appropriate thin BP flake. A bright field image of a BP flake on a PDMS film is shown in Fig. 1(b). The image was taken using a home-made optical microscope equipped with an objective lens (numerical aperture, N.A., is 0.45) and measured under slightly-oblique white light illumination through the lens. We can easily identify atomically-thin BP regions (indicated by the white dash line) through the color and contrast of the image. We further characterize the BP flake by PL experiments using the same microscope. The PL measurements were performed at room temperature under a continuous wave (cw) laser pumping at 785 nm through the objective lens. The same lens collects and guides the PL signal, which was analyzed by a spectrometer equipped with an InGaAs array detector. Figure 1(c) shows a PL spectrum of the atomically-thin BP shown in Fig. 1(b), measured under a pump power of 1 mW (measured before the objective lens). Bright PL signal centered around 1075 nm indicates that the BP region flake is composed of 2 atomic layers. Within our detection wavelength range (900-1600 nm), we observe three different types of luminescent BPs respectively centered around 1100, 1450, >1600 nm, corresponding to the discrete increase of number of layers as observed in the literature[19]. It is noteworthy that we did not observe any degradation of PL signal during the sample search process in the vacuum chamber kept under a pressure of < $10^{-3}$ Pa; in contrast, noticeable degradation was observed when optically characterizing BP flakes in the glove box filled with dry nitrogen gas.

2D PhC nanocavities were prepared on a GaAs substrate in an airbridge form by standard semiconductor nanofabrication. We define the defect nanocavity by introducing three missing air holes in a triangular PhC air hole lattice (hole radius= 79 nm, period = 310 nm) patterned on a 130-nm-thick GaAs slab (refractive index, *n*, is 3.4). Using finite difference time domain simulations, we calculated a field profile of the fundamental cavity mode resonating around $\lambda$ = 1090 nm, as shown in Fig. 1(d). A cross section of the field distribution is exhibited in Fig. 1(e), confirming the existence of strong evanescent optical field on the surface of the nanocavity: the field amplitude here is 70% to the field maximum. The positions of air holes near the defect regions are systematically shifted for achieving a better optical confinement[22]. We also introduce a double-period modulation of the air hole size in order to increase the light out-coupling efficiency[23]. As a result, the cavity mode possesses a calculated *Q* factor of 7,000 while keeping a small mode volume of 0.82 $(\lambda/n)^3$. Using cross polarized reflectivity measurements[24] at room temperature, we experimentally confirmed high *Q* factors over 5,000 for the PhC nanocavities without BPs on top of them (See the supplemental material for the spectrum). All of these cavity properties are advantageous for achieving a strong light matter interaction.

Finally, we transfer the BP flake after the PL measurement to an airbridge PhC nanocavity under a home-made transfer printing apparatus that is built in a glove box filled with nitrogen gas. We put the atomically-thin region of the BP flake accurately on top of the nanocavity under a microscope equipped with precision position control stages. The optical microscope employs a NIR illumination light source for mitigating the erosion of BP flakes. After putting the BP flake, we slowly peeled off the PDMS stamp, solely leaving the BP flake on the cavity surface, as confirmed in an optical microscope image shown in Fig. 2(a). The BP flake is placed close to the defect cavity region, although the atomically-thin BP region is not anymore clearly visible. Figure 2(b) shows a scanning electron microscope image of the fabricated sample (taken after all the measurements performed in this work). It is clearly seen that the cavity defect is covered with the thin BP, which seems to be torn around a boundary across the defect cavity. In our fabrication processes, we often observe this phenomenon, which could be useful for selective transfer of atomically-thin BP regions to photonic nanostructures. We also found that the success probability of the transfer process largely depends on the surface conditions of the GaAs slab: the process is more likely to succeed when being carried out just after removing the surface oxide on GaAs PhC cavities.

Figure 2(c) shows a comparison of PL emission intensities for the atomically-thin BP before and after the transfer process, measured using the same PL setup for searching the BP flake. The PL is taken near the edge of the atomically-thin region, which, for the case after the transfer, is placed on an unpatterned region of the GaAs substrate. The center wavelength of the PL is 1107 nm, which is shifted to a longer wavelength from that measured on PDMS, probably originating from to the difference in the electric permittivities of the two substrate materials[25]. The PL signal after the transfer appears weaker than that before the transfer (a 60 % reduction of the integrated intensity), although such simple comparison should not be justified since the efficiencies of the excitation and output collection should be different between the BP on PDMS and on GaAs. Nevertheless, the relatively-small reduction of the integrated PL intensity can be regarded as an indication that our transfer process prevents the degradation of BP flakes to a certain level.

Next, we characterize the optical coupling between the atomically thin BP and the nanocavity. For this experiment, we used another micro PL setup equipped with an objective lens with a higher numerical aperture of 0.65. We pump the defect cavity region of the fabricated BP-nanocavity system by a 780 nm cw laser with a power of 50 μW at room temperature. A measured PL spectrum is shown in Fig. 3(a). Besides the broad emission peak originated from the two layer BP, a sharp resonance peak is observed at 1074 nm. Figure 3(b) shows a higher resolution spectrum for the sharp resonance peak, exhibiting a narrow linewidth of ~ 4 nm, corresponding to a quality factor of ~ 260. This observed degradation of $Q$ factor compared to that without BP could be mainly attributed to strong optical absorption in BP, although further detailed studies are necessary to verify it. We consider that the observed sharp resonance peak originates from the coupling of BP emission into the fundamental cavity mode. Indeed, the resonance wavelength is close to those measured for bare PhC nanocavities by reflectivity experiments (~ 1070 nm). The direction of linear polarization of the cavity mode (not shown) is also consistent with that of the cavity mode. We also note that bare PhC cavities without the coverage of BPs does not exhibit any PL signal under the same PL measurement condition, suggesting that there are no noticeable light source other than the thin BP flake in our sample. In order to further confirm the observation of cavity mode emission originated from the BP emission, we characterized the position dependence of PL signal, as shown in Fig. 3(c) and (d). We found that the cavity emission intensity is well localized around the defect cavity region. The PL intensities spread relatively widely in X direction, in which the defect cavity elongates, while stronger localizations can be seen in Y direction. We also measured polarization properties of the fabricated sample with a set of a half waveplate and a linear polarizer inserted in the signal detection path. Figure 3(e) shows integrated peak intensities of the cavity (red) and BP (blue) emission plotted as a function of half waveplate angle. Both emission peaks exhibit sinusoidal curves, confirming their linearly polarized emission. We found that the optical axes of the BP and the cavity mode deviate by 60 degree each other, confirming that the polarization of the BP emission coupled to the cavity mode is governed by that of the cavity mode. The use of the linear polarizer is helpful to distinguish the portion of BP emission that couples to the cavity mode from that does not. Overall, these observations further supports that the sharp emission peak in Fig. 3(a) and (b) originates from BP emission coupled to the PhC cavity mode, as a result of their mutual optical coupling. We note that, in order to further evaluate the strength of coupling between the BP flake and the cavity mode, it would be useful to study PL emission dynamics with time-resolved measurements.

It is noteworthy that we did not observe laser oscillation from any of the several samples fabricated in the same manner. They are measured under strong cw and pulsed (pico second pulse at 780 or 920 nm) pumping conditions and also under low temperatures from 10 K to 150 K. Some of the samples were fabricated using nanocavities with much higher design $Q$ factors over 50,000. We also examined a few samples under the resonance condition between the BP emission center wavelength and the cavity mode, however they did not lase. One possible explanation of these observations is simply the lack of sufficient material gain in the atomically-thin BPs for lasing in the current device setup. Another view point would be insufficient cavity quality for supporting the laser oscillation. In our experiments, the cavity $Q$ factor may have be degraded under the dense carrier injection that induces free carrier absorption of the intracavity photons. We expect future experimental and theoretical works to clarify the possibility of NIR lasing using atomically-thin BPs.

In summary, we demonstrated optical coupling between an atomically-thin BP and a PhC defect nanocavity. We fabricated a BP-on-nanocavity structure using dry transfer technique implemented in an inert gas atmosphere that mitigates the notoriously rapid degradation of thin BP flakes in air. We observed cavity coupled emission from the 2D BP in the NIR with a sharp resonance having a $Q$ factor of 260. The optical coupling between the BP and the cavity is further confirmed through measuring the position dependence of the emission intensity. We believe that our results are an important step toward the development of 2D material based nanophotonic devices efficiently operating in the NIR, including those behaving as light sources like nanolasers.

SUPPLEMENTARY MATERIAL

See supplementary material for the cavity resonance spectrum without BP taken by the cross polarized reflectivity measurement.

ACKNOWLEDGEMENTS

The authors thank C. F. Fong for fruitful discussions. This work was supported by JSPS KAKENHI Grant-in-Aid for Specially Promoted Research (15H05700), JSPS KAKENHI Grant Numbers JP16K06294, JP15K17433, JP16H00982, JP25107003, and JP26248061, Murata Scientific Foundation, NEDO project, and JST CREST Grant Numbers JPMJCR15F3.

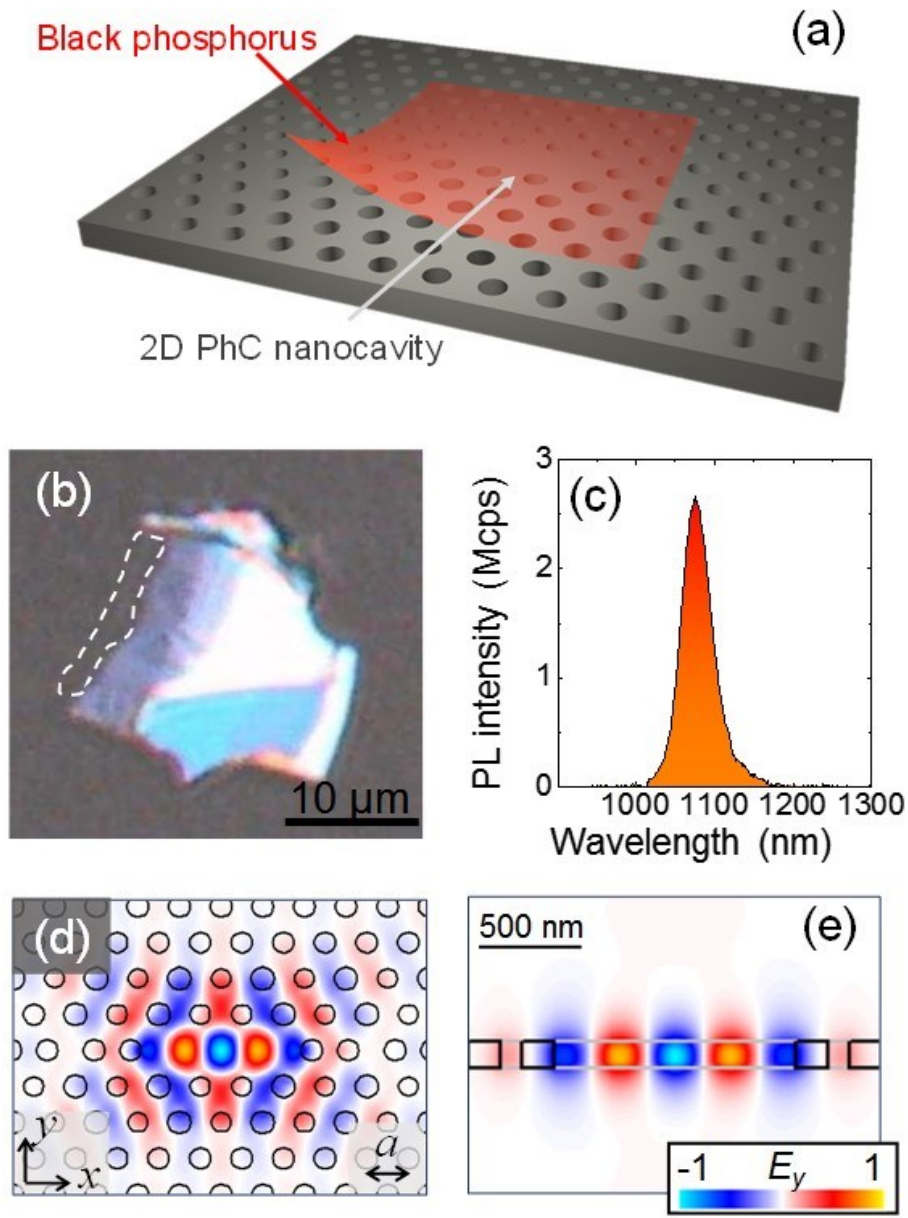

Figure 1. (a) Schematic illustrations of an atomically-thin BP film placed on a 2D PhC nanocavity. The edge of the BP is pulled up for the clarity. (b) Optical microscope image of a BP flake prepared on a PDMS film. The region highlighted by the white dash line is composed of two atomic BP layers. (c) PL spectrum of the two layer BP region, exhibiting a bright luminescence peaked at 1074 nm. (d) 2D PhC nanocavity design, overlaid with a simulated field profile for the fundamental cavity mode. (e) Cross section of the cavity field profile, indicating the existing of strong optical field on the surface of the PhC slab.

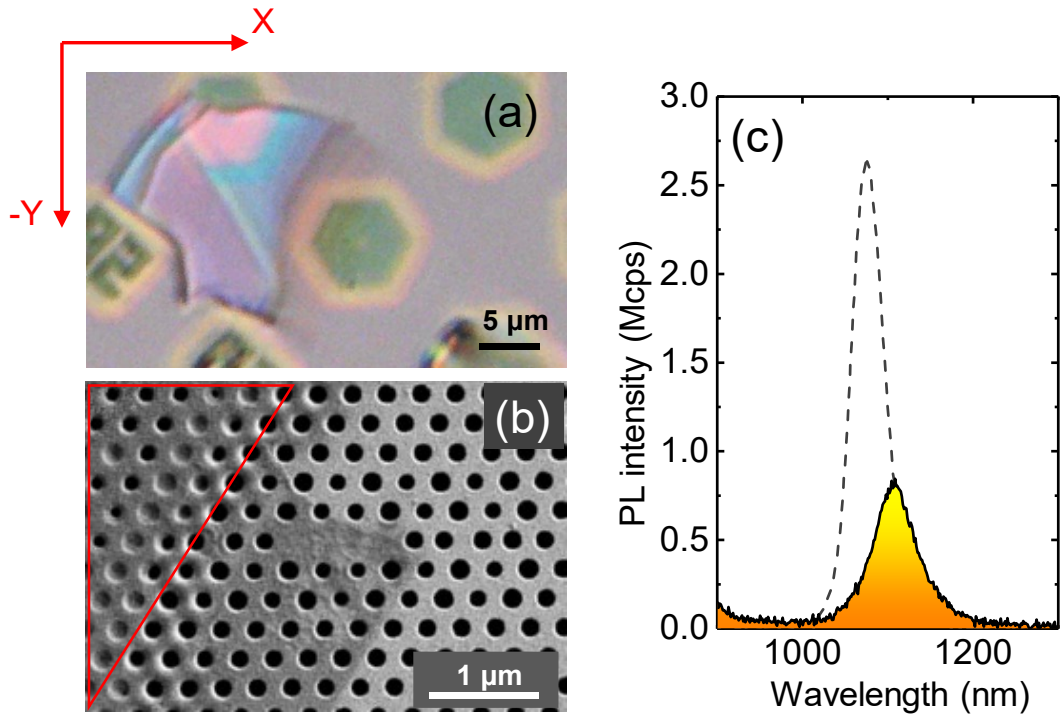

Figure 2. (a) Optical microscope of the fabricated BP-on-nanocavity sample. (b) Scanning electron microscope image of the same sample, taken after all the measurements conducted in this work. Regions with darker contrasts are covered with BP. A region enclosed in the red line in the picture is composed of thicker BP layers, deduced from a comparison with optical microscope images. (c) Comparison of PL emission intensities between before (dashed gray line) and after (solid line, yellow shaded) the transfer. The PL spectrum after the transfer was taken near the edge of the atomically-thin region, which is placed on an unpatterned region of the GaAs substrate.

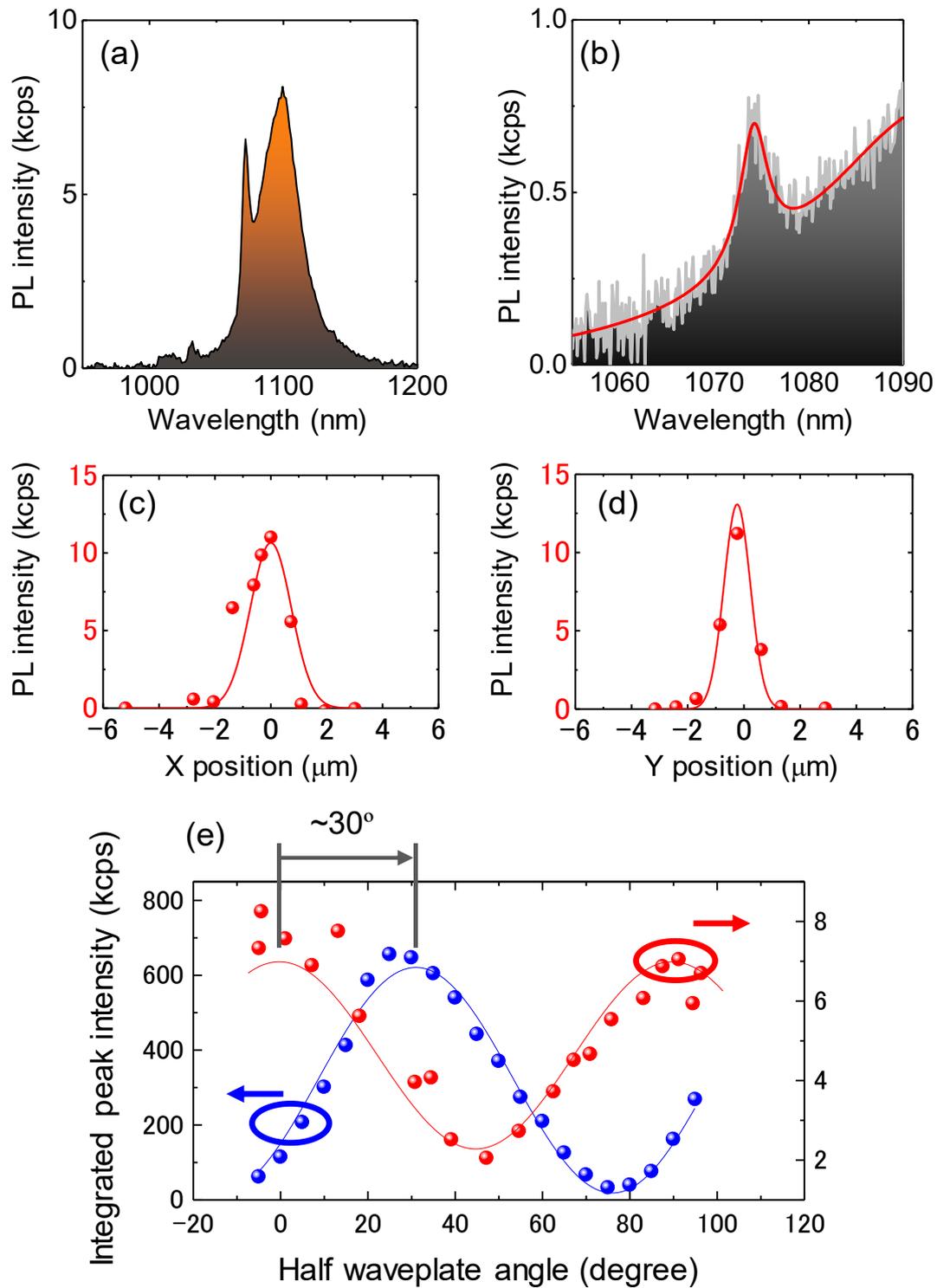

Figure 3. (a) PL spectrum taken under the excitation of the defect cavity region of the fabricated sample. Broad emission peak centered at 1100 nm is from the 2D BP, while the sharp peak at 1070 nm is from the BP emission coupled to the fundamental cavity mode. (b) High resolution PL

spectrum of the cavity mode emission. A cavity *Q* factor of 260 can be extracted through a Lorentzian peak fitting (red solid curve) taking into account the background BP emission. (c) Excitation position dependence of integrated PL intensities of the cavity mode emission measured along x direction. The integrated PL intensities are deduced from multi-peak fitting. The axis is defined in Fig. 2(a). (d) Same in (c) but along y direction. (e) Integrated peak intensities of the cavity mode (red) and BP (blue) emission, plotted as a function of half waveplate angle. The phases of the sine oscillations deviate ~ 30 degree each other, corresponding to the deviation of the two optical axes of ~ 60 degree.

# Supplementary material for: Optical coupling between atomically-thin black phosphorus and a two dimensional photonic crystal nanocavity

Yasutomo Ota[1], Rai Moriya[2], Naoto Yabuki[2], Miho Arai[2], Masahiro Kakuda[1],
Satoshi Iwamoto[1,2], Tomoki Machida[1,2] and Yasuhiko Arakawa[1,2]
E-mail: ota@iis.u-tokyo.ac.jp

*1) Institute for Nano Quantum Information Electronics, The University of Tokyo, 4-6-1 Komaba, Meguro-ku, Tokyo 153-8505, Japan*

*2) Institute of Industrial Science, The University of Tokyo, 4-6-1 Komaba, Meguro-ku, Tokyo 153-8505, Japan*

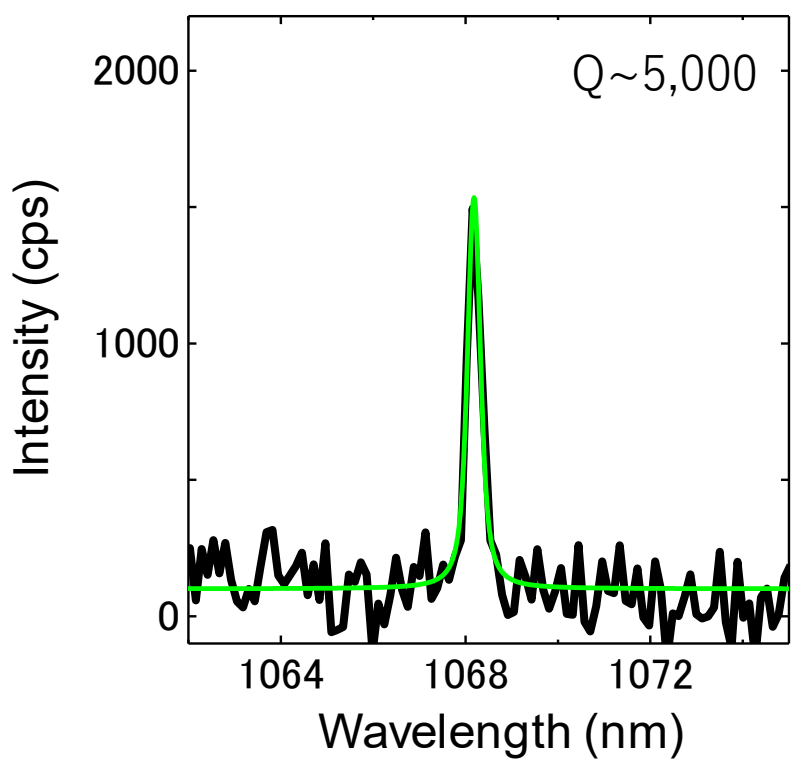

Figure S1. Resonance spectrum for a typical bare PhC nanocavity. The spectrum is measured using the cross-polarized reflectivity setup.